\documentclass[aps,prl,epsfig,showpacs,superscriptaddress,twocolumn]{revtex4}

\usepackage{graphicx,amsmath}

\begin{document}

\title{Domain wall magnetoresistance in a nanopatterned La$_{2/3}$Sr$_{1/3}$MnO$_3$ track}

\author{T. Arnal}
\affiliation{Institut d'Electronique Fondamentale, CNRS, Universit\'e Paris-Sud, 91405 Orsay, France}

\author{A.V. Khvalokvskii}
\address{General Physics Institute of the Russian Academy of Science, Vavliova Str. 38, 119991 Moscow, Russia}

\author{M. Bibes}
\affiliation{Institut d'Electronique Fondamentale, CNRS, Universit\'e Paris-Sud, 91405 Orsay, France}

\author{Ph. Lecoeur}
\affiliation{Institut d'Electronique Fondamentale, CNRS, Universit\'e Paris-Sud, 91405 Orsay, France}

\author{A.-M. Haghiri-Gosnet}
\affiliation{Institut d'Electronique Fondamentale, CNRS, Universit\'e Paris-Sud, 91405 Orsay, France}

\author{B. Mercey}
\affiliation{CRSIMAT-ISMRA, 6 Bd. du Mar\'echal Juin, 14050 Caen cedex, France}

\date{\today}

\begin{abstract}


We have measured the contribution of magnetic domain walls (DWs) to the electric resistance in epitaxial
manganite films patterned by electron-beam lithography into a track containing a set of notches. We find a DW
resistance-area (RA) product of $\sim$2.5 10$^{-13}$ $\Omega$m$^2$ at low temperature and bias, which is several
orders of magnitude larger than the values reported for 3d ferromagnets. However, the current-voltage
characteristics are highly linear which indicates that the DWs are not phase separated but metallic. The DWRA is
found to increase upon increasing the injected current, presumably reflecting some deformation of the wall by
spin-transfer. When increasing temperature, the DWRA vanishes at $\sim$225K which is likely related to the
temperature dependence of the film anisotropy.

\end{abstract}

\maketitle

The influence of magnetic domain walls (DWs) on the electric resistance of ferromagnets has been the subject of
numerous studies in the past \cite{kent2001}. At present, DWs are still intensely investigated as they are
expected to play an important role in future spintronics devices. For example, it has been shown that DWs can be
displaced by spin-polarized currents, which can be used to switch the magnetic configuration of a magnetic
trilayer from parallel to antiparallel \cite{ono98,grollier2003}. Even more, logic circuits based on magnetic
DWs have been proposed recently \cite{allwood2005}.

Most experimental works on the transport properties of magnetic DWs have focused on conventional 3d ferromagnets
\cite{kent2001}. In these systems, the DW magnetoresistance (DWMR, defined as the change in resistance between
the case where the sample is in a single domain state, with all magnetic moments aligned, and the case where the
sample is divided into domains separated by domain walls that electrons must cross) is usually less than 1\%,
with DW resistance-area (RA) products in the 10$^{-19}$-10$^{-16}$ $\Omega$m$^2$ range \cite{kent2001}. More
recently, the DWMR has also been measured in (Ga,Mn)As \cite{ruster2003} and in ferromagnetic oxides like
SrRuO$_3$ \cite{klein2000} and manganites \cite{mathur99,li2001,wolfman2001,cespedes2005,pallecchi2006}. In
manganese perovskites especially, the strong interplay between the spin, charge and lattice degrees of freedom
\cite{coey99} is expected to provide an additional richness to the physics of electron transport through DWs
\cite{mathur2001,golosov2003,rzchowski2004}.

Little is known on magnetic DWs in manganites. Fresnel imaging has been used to measure the width $\delta$ of a
DW in a 200 nm La$_{2/3}$Ca$_{1/3}$MnO$_3$ (LCMO) film, yielding $\delta\simeq$ 38 nm \cite{lloyd2001}. Their
contribution to the resistance was measured in patterned LCMO tracks and a RA product of 8 10$^{-14}$
$\Omega$m$^2$ was found at 77K \cite{mathur99}, which is several orders of magnitude larger than what is found
in conventional 3d metals. Estimations based on a simple application of the double-exchange model revealed that
the resistivity increase due to conventional Bloch-type DWs cannot possibly explain experimental results
\cite{yamanaka96,mathur99}. To account for this discrepancy, suggestions that DWs in manganites have an unusual,
non-Bloch structure were made \cite{mathur2001}. In reference \onlinecite{mathur2001} it was predicted that for
manganites with intermediate doping level two types of DW may take place. In most of the cases the
long-wavelength properties of double-exchange ferromagnets are adequately captured within an effective
Heisenberg description and conventional Bloch-type DWs are a stable solution of the corresponding Hamiltonian
\cite{golosov2000}. However under specific conditions stripe walls, which are composed of two ferromagnetic
domains separated by a stripe of antiferromagnetic phase (phase-separated DW), may become more energetically
favourable than Bloch-type walls. Provided that such DWs exist, their contribution to the total resistance of
the sample should be substantially large, as carrier transport across a phase-separated DW is strongly
suppressed due to insulating properties of the antiferromagnetic phase. It was suggested that these
phase-separated DW may account for the large DW resistance observed in previous experiments
\cite{mathur2001,golosov2003}. However no clear evidence of this fact exists.

Here, we report magnetotransport measurements of devices consisting of a track with nanometric notches defined
in a La$_{2/3}$Sr$_{1/3}$MnO$_3$ (LSMO) epitaxial thin film. In the R(H) dependence of these devices, we observe
jumps in the resistance at low magnetic field (H) which we attribute to domain walls. Their RA product is in the
10$^{-13}$ $\Omega$m$^2$ range and, remarkably, increases with the applied bias. We argue that in our case the
DWs are not of the phase-separated type and should be depinned by critical currents substantially lower than in
3d metals.

LSMO thin films (thickness t = 10 nm) were grown on (001)-oriented SrTiO$_3$ (STO) substrates by pulsed laser
deposition \cite{lecoeur2006}. The films are epitaxial as evidenced by X-ray diffraction studies, and their
surface shows flat terraces separated by one-unit-cell-high steps. The grain size is typically $\sim$100 $\mu$m.
The Curie temperature (T$_C$) of the films was 320 K, i.e. slightly lower than the bulk T$_C$ (360K) due to
their low thickness \cite{maurice2002}. M(H) cycles measured at 10K along different in-plane directions
evidenced a biaxial anisotropy, with the easy axes along [110] and [-110] (Ref. \onlinecite{arnal2006}). The
corresponding cubic anisotropy constant was K=9 10$^3$ Jm$^{-3}$.

\begin{figure}
\includegraphics[width=0.9\columnwidth]{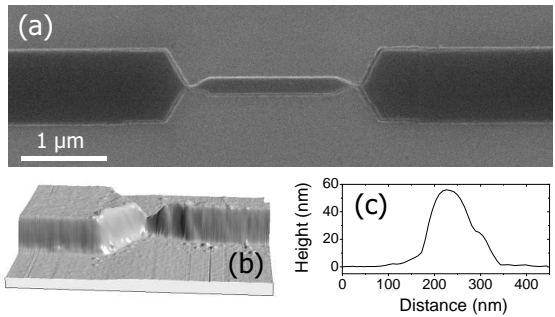}
\caption{(a) Scanning electron microscopy image of the device, i.e. the LSMO track covered by the resist, after
ion beam etching. (b) AFM image of one of the notches. (c) Profile along the dashed line in (b).} \label{afm}
\end{figure}

The films were patterned by electron-beam lithography, using a negative resist HSQ-FOX 12. No subsequent
annealing was applied. The exposition was performed in a modified field-effect gun scanning electron microscopy
(SEM) at 30 keV (see Ref. \onlinecite{arnal2006} for details). After exposure, development and ion-beam etching,
the $\sim$100 nm thick (insulating) resist was left on the patterned track, although its thickness was strongly
reduced at the notches. The devices defined with this process consist of a 350 nm wide and 2 $\mu$m long central
domain connected on each side to 1 $\mu$m wide and 15 $\mu$m long tracks by two narrow notches. The track was
oriented along the easy [110] direction. The dimensions of the devices were determined in view of micromagnetic
simulations results and chosen so as to allow magnetization reversal in the central region at a larger field
than in the outer arms. Thus, in a certain field range corresponding to an antiparallel alignment of the central
domain magnetization with respect to the magnetization in the outer arms, a DW should be located at each notch.
Micronic sputtered Au pads were used as contacts. Figure \ref{afm}a show a SEM image of one of the devices.
Atomic force microscopy (AFM) images (Fig. \ref{afm}b) and cross-sections (Fig. \ref{afm}c) of the notches
revealed that their width is around w=150 nm.

\begin{figure}
\includegraphics[width=0.9\columnwidth]{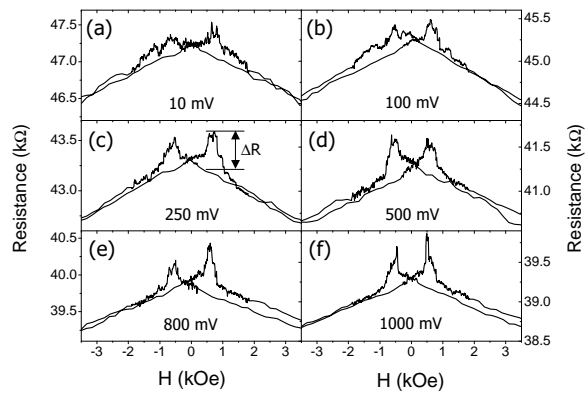}
\caption{R(H) curves at 8K and different bias voltages.} \label{mr}
\end{figure}

In figure \ref{mr}, we show a set of R(H) curves measured at 8K with different applied bias voltages (V$_{DC}$)
for the device of figure 1. When coming from high positive magnetic field to low negative H values a jump
$\Delta$R occurs in the resistance (between -100 and -500 Oe). Upon increasing the field further to large
negative values, R shows some sharp features and then recovers a low-resistance state at about -1 kOe. A
virtually symmetric behaviour is obtained when going back to large positive fields. These jumps are not observed
in a track without notches \cite{arnal2006}. We also note that at large field, the resistance decreases
linearly, which reflects some spin disorder.

We have verified that the typical artifacts that occur in DWMR measurements \cite{kent2001} cannot explain our
data. We have measured the MR of a track without notches with the field applied in plane and perpendicular to
the track direction. The R(H) is dominated by a linear negative MR in all the field range, as observed at high
field in the device R(H) data of figure \ref{mr}. This indicates that the contribution of AMR is very small (see
inset of Fig. \ref{temp}). As our films are magnetized in plane, we can also discard the contribution of Hall
effect. Finally, the Lorentz MR is negligible in manganites due to their very small carrier mean free path
($\lambda\simeq$ 1 nm). We thus conclude that the jumps observed in the R(H) cycles most likely arise from the
presence of DWs. We can calculate the resistance-area product (RA) for the DWs, assuming they are located at the
notches. From the value of $\Delta$R and taking a sectional area of 150$\times$10 nm$^2$, we find that a DW has
an RA of $\sim$2.5 10$^{-13}$ $\Omega$m$^2$ at 8K and $|$V$_{DC}|$=200 mV. This is in good agreement with the
value reported by Mathur et al \cite{mathur99}, larger by several orders of magnitude than what is found for 3d
metals, and about 100 times larger than the DWRA in SrRuO$_3$ \cite{klein2000}. Our results thus confirm that
DWs in manganites are highly resistive, with RA values several orders of magnitude larger than what is expected
from a simple double-exchange model \cite{mathur99}.

\begin{figure}
\includegraphics[width=0.9\columnwidth]{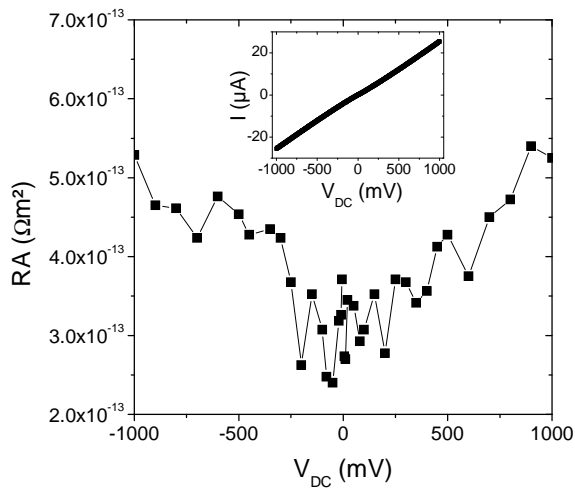}
\caption{Bias voltage dependence of the DWRA, at 8K. Inset : I(V) curve of the device.} \label{bias}
\end{figure}

Further insight on the nature of the DW can be gained from inspecting the I(V) curves of the device. As visible
on the inset of figure \ref{bias}, they are highly linear, with only a small non-linear contribution. This rules
out the presence of insulating regions in the track such as grain boundaries or charge-ordered antiferromagnetic
stripes that would act as tunnel barriers and thus lead to highly non-linear I(V) curves. We thus conclude that
our DWs are not of the phase-separated type \cite{mathur2001,golosov2003}. The small non-linear contribution is
virtually independent of magnetic field and thus probably reflects the existence of localized states related to
structural defects created by the lithography process.

Since our data do not support that phase-separated DWs are responsible for the huge resistance of DWs in LSMO,
alternative scenarios must be found. It is known that narrow-band manganites can show a metallic behavior at low
temperature and have resistivities of up to $\sim$10$^5$ $\Omega$cm, i.e. 8 orders of magnitude larger than in
LSMO \cite{coey95}. While the reduction of the double-exchange interaction due to the rotation of moments within
the DW is not enough to explain this large DWRA \cite{mathur99}, it is possible that within the DW, the
conduction band becomes more narrow due the reinforced competition between a weakened double-exchange
interaction and super exchange. This may lead to a reduced carrier mobility or to the trapping of some carriers
as is thought to occur in narrow-band manganites \cite{coey95}, hence increasing the resistivity.

The thick resist layer left on the patterned track impeded its magnetic imaging. To get a better insight into
the micromagnetics of our device, we performed simulations. This study revealed that the DWs pinned at the
notches should be of the head-to-head type. As the films are intrinsically very soft, notches may constrain
\cite{bruno99} the DWs. The high switching fields found experimentally evidence a significant role of structural
defects in the reversal processes. These defects may result in further constriction of the DWs. Details of this
study will be given elsewhere \cite{khvalkovski2006}.

The main panel of figure 3 shows the variation of the wall resistance with bias voltage. Remarkably, the
amplitude of DWRA increases upon increasing the bias voltage (in absolute value) in the -1V to 1V range, see
figure 3. This behaviour is qualitatively very different from that observed in manganite-based tunnel junctions,
in which the tunnel magnetoresistance (TMR) is found to decrease with bias, under the influence of
electron-magnon scattering and band structure effects \cite{bowen2005}. This further precludes the existence of
a charge-ordered antiferromagnetic core in the DWs. Note that the variation of DWRA we observe here is unlikely
to be caused by Joule heating, as the device resistance decreases with increasing bias voltage (see figure 2)
while it increases with temperature \cite{arnal2006}. A possible explanation is the deformation of the DW under
the influence of the injected current, that is the transfer of spin angular momentum from the spin-polarized
current to the local magnetic moment. This spin-pressure effect has been predicted theoretically and
experimentally observed at current densities of J$\sim$10$^{12-13}$ Am$^{-2}$ in NiFe structures
\cite{kimura2003b,klaui2005}. In our case, J is much smaller, on the order of 10$^{10}$ Am$^{-2}$, but the
current spin-polarization in LSMO is almost 100\% (Ref. \onlinecite{bowen2003}) and, as previously mentioned,
the DWs are very resistive, so that even a small deformation should produce a visible change of DWRA. We also
point out that this observation suggests rather low critical current densities for domain wall depinning and
motion in LSMO, as also found by Pallecchi \emph{et al} \cite{pallecchi2006b}.

\begin{figure}
\includegraphics[width=0.9\columnwidth]{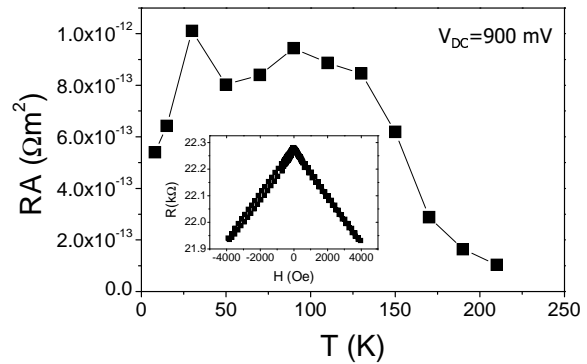}
\caption{Temperature dependence of the DWRA. Inset : R(H) of a track without notches at 8K, 100 mV and with the
field applied in-plane and perpendicular to the track direction.} \label{temp}
\end{figure}

We have also measured the temperature dependence of the DW resistance, see figure \ref{temp}. $\Delta$R is
almost constant up to about 125K and then decreases to vanish around 225K, which is substantially lower than the
T$_C$ of the film (300K). In manganites, extrinsic magnetoresistive effects such as powder magnetoresistance
\cite{hwang96} or TMR \cite{garcia2004} are known to decrease with temperature and disappear at temperatures
lower than T$_C$. In the case of TMR, this is ascribed to a reduced T$_C$ at the interfaces between the
manganite electrodes and the barrier. Here, a possibility would also be that the manganite in the notches
regions has a depressed T$_C$ compared to the rest of the film, due to size effects or deterioration by the
fabrication process. However in that case, a signature of the metal-insulator transition (occurring at T$_C$ in
manganites \cite{coey99}) of the notches should have been observed in the R(T) curve \cite{arnal2006}. An
alternative explanation resides in the temperature dependence of the biaxial anisotropy constant in LSMO that
decreases rapidly with T and vanishes close to T$_C$ \cite{steenbeck99}. Therefore, the DW is expected to
broaden when T increases, which should reduce its contribution to resistance. Temperature dependent magnetic
imaging should bring insight on this point.

In summary, we have measured the resistance of magnetic domain walls in a LSMO track containing two $\sim$150 nm
wide notches. We find a DWRA of 2.5 10$^{-13}$ $\Omega$m$^2$ at low temperature and bias, which increases when
increasing the injected current, possibly reflecting the DW deformation by spin-transfer. I(V) curves are almost
linear which indicates that the DWs in LSMO are not of the phase-separated type \cite{golosov2003}. The DWMR is
found to vanish at about 225K, which we tentatively ascribe to the broadening of the DW due to the temperature
dependence of the anisotropy constant of LSMO. The next step would now consist in studying the current-induced
DW motion in half-metallic manganites. Indeed, the large DW resistivity should make LSMO well suited for testing
DW-based logic architectures with resistive readout.

\vspace{0.5cm}

\acknowledgments{Financial support by the French ACI Nanomemox, the RFBR project 04-02-17600 and ECONET is
acknowledged. We thank B. Montigny for the AFM images, A. Aassime for his help with the lithography process and
K.A. Zvezdin, A.K. Zvezdin and D. S\'anchez for fruitful discussions.}


\end{document}